\documentclass[11pt]{article}
\usepackage{amsmath,amssymb,bm}
\usepackage{graphics}
\usepackage{graphicx}
\usepackage{amssymb}
\usepackage{epstopdf}

\usepackage{mystyle}
\usepackage{cite,./mcite}
\usepackage{authblk}
\usepackage{lineno}
\usepackage{wrapfig}

                \def\lsim{\mathrel{\rlap{\lower4pt\hbox{\hskip1pt$\sim$}}
    \raise1pt\hbox{$<$}}}                \def\gsim{\mathrel{\rlap{\lower4pt\hbox{\hskip1pt$\sim$}}
    \raise1pt\hbox{$>$}}}

\begin{document}

\title{A Brief Comment on Multi-Gluon Amplitudes and Double Parton Interactions}
\author{Daniele Treleani}

\author[1]{Giorgio Calucci}
{\tiny
\affil[1]{University of Trieste and INFN}
} 
\date{ }
\maketitle

\begin{abstract}
A typical contribution to a color ordered multi-gluon amplitude, which can split into two weakly correlated two-body gluon scattering amplitudes and may thus contribute to a Double Parton Interaction, is briefly discussed. We find that the color ordered amplitude is not enhanced in the typical configuration generated by a DPI, where the transverse momenta of final state gluons are compensated pairwise, while a dominant contribution to the multi-gluon amplitude is due to terms proportional to the fusion amplitude of two initial state gluons. Which corresponds to an amplitude effectively describing a two rather than a three-body partonic interaction.
\end{abstract}

\begin{wrapfigure}{r}{0.5\textwidth}
  \vspace{-25pt}
  \begin{center}
    \includegraphics[width=0.6\textwidth]{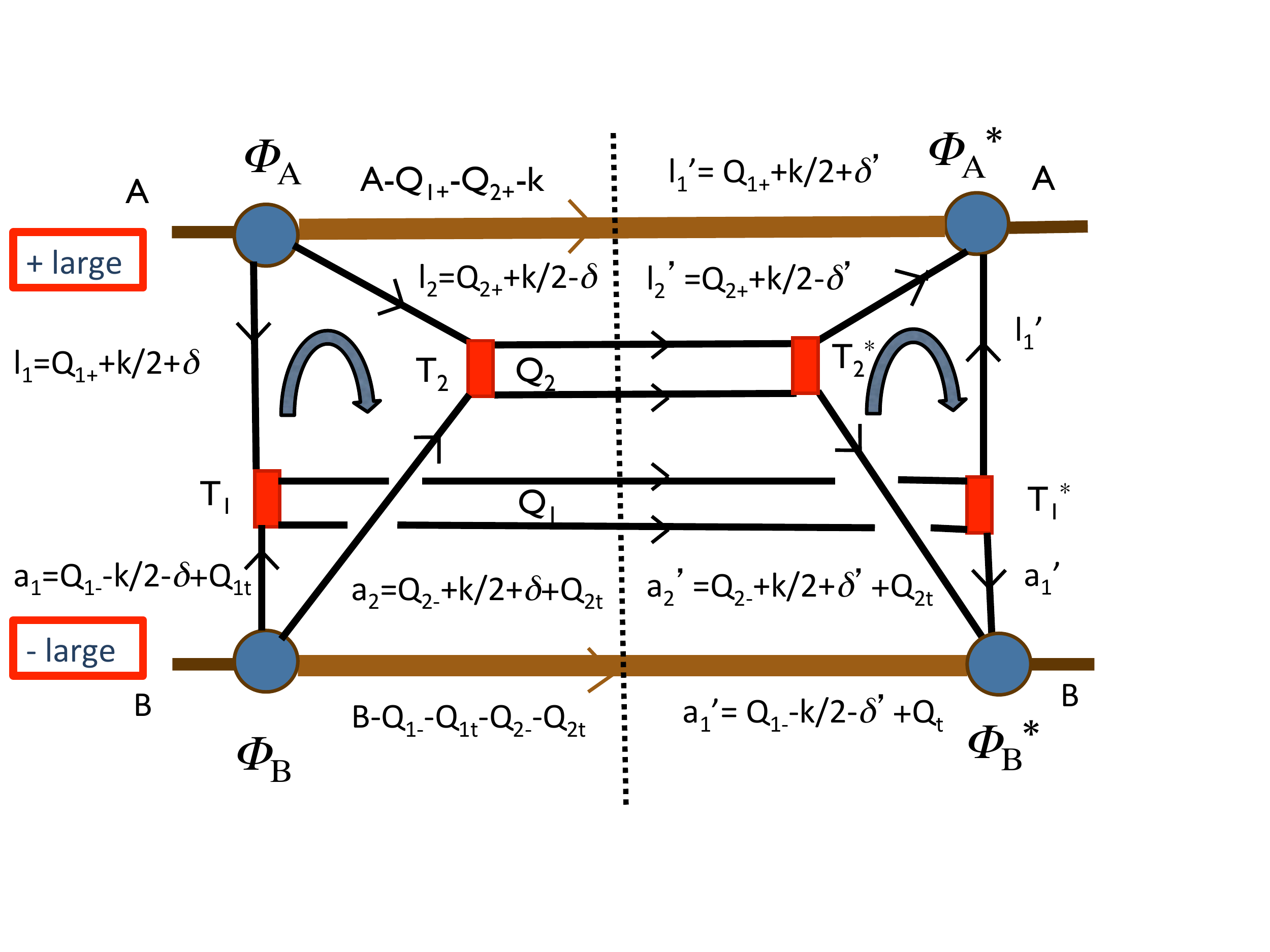}
  \end{center}
  \vspace{-30pt}
  \caption{\small Double parton scattering contribution to the forward elastic amplitude}
  \vspace{-5pt}
\end{wrapfigure}

 The Double Parton Interaction cross section can be expressed as a contribution to the forward elastic scattering amplitude characterized by two independent loops (cfr. Fig.1), where the initial momenta of the partonic interactions are integrated independently in the amplitude and in its complex conjugate, is such a way that the process is not diagonal as a function of the initial parton's momenta \cite{Paver:1982yp}. To evaluate the loop integral, it is convenient to distinguish two qualitatively different terms in the vertex ${\it \Phi}$, representing the hadron structure (cfr. Fig.2). Namely the short distance contribution, $\phi$, and the long distance contribution, $\psi$. In this way the interaction amplitude splits into 4 different terms. As shown in Fig.3, the first 3 terms contribute to DPIs, since the vertex $\psi$ introduces a non perturbative dimensional factor in the amplitude, while the 4th term, where only $\phi$ vertices are present, contributes as a loop correction to the $2\to4$ parton scattering amplitude\cite{Blok:2011bu}\cite{Diehl:2011yj}.

The non perturbative dimension, characterizing the vertices $\psi_A$ and $\psi_B$, limits the virtualities of the partons in the the loop to values of the order of the hadronic mass. As a consequence, the upper part of the loop in the interaction amplitude is characterized by large light cone '$+$' components, while the lower part of the loop is characterized by large light cone '$-$' components, in such a way that the loop integrations on the light cone '$+$' and '$-$' integration variables can be done independently. 

\begin{figure}
 \vspace{-40pt}
\begin{minipage}[b]{8cm}
\centering
\includegraphics[width=8.5cm]{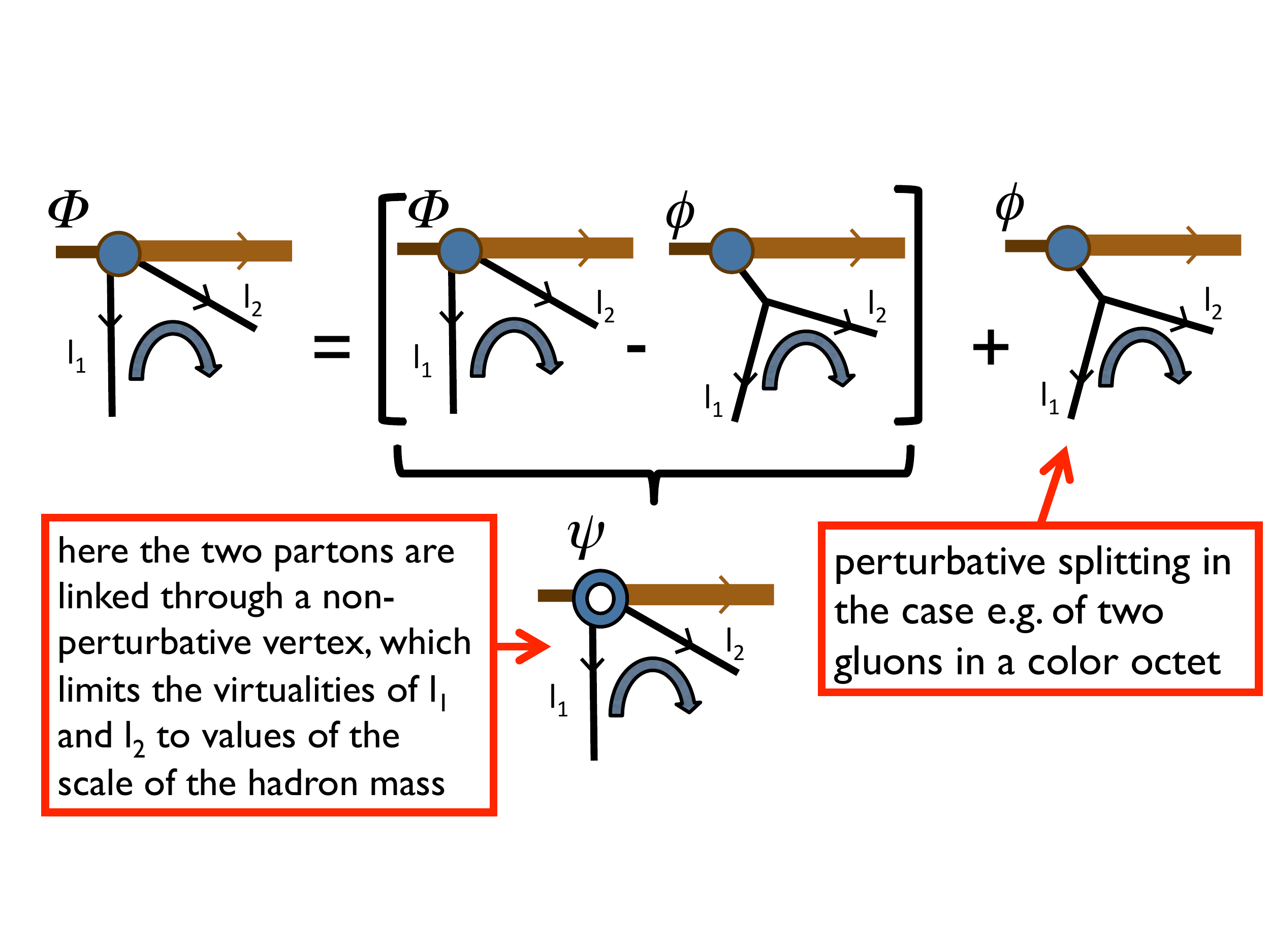}
\vspace{-25pt}
\caption{\small Non perturbative and perturbative contributions to the double parton vertex}
\end{minipage}
\ \hspace{3mm} 
 \vspace{-40pt}
\begin{minipage}[b]{8cm}
\centering
\includegraphics[width=8.8cm]{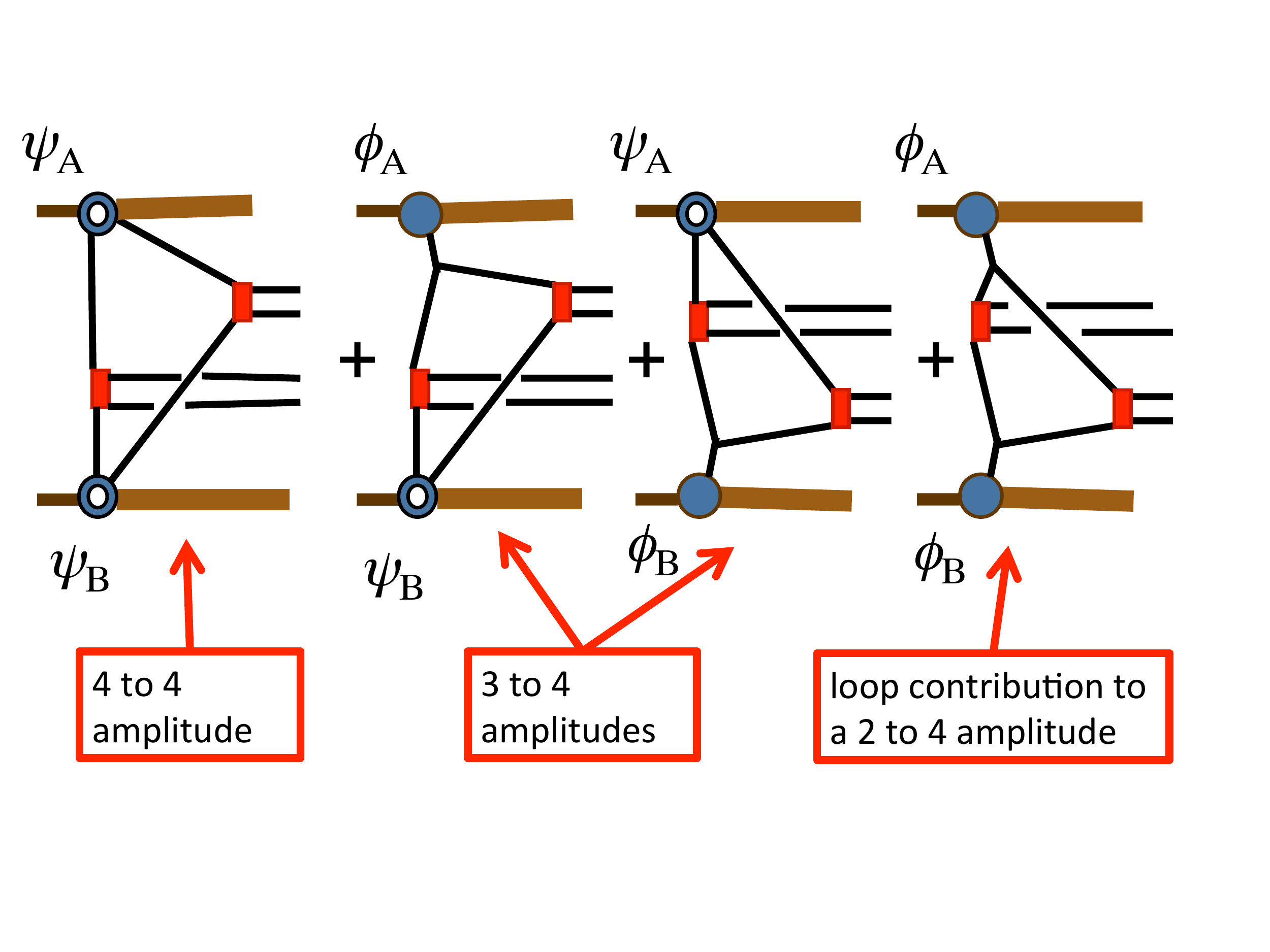}
\vspace{-25pt}
\caption{\small Different $4\to4$, $3\to4$ and $2\to4$ contributions to the interaction amplitude}
\end{minipage}
\vspace{30pt}
\end{figure}

In the case of the $4\to4$ scattering process, the feature is illustrated in Eq.(1) where the integrations on the loop variables $\delta_-$ and $\delta_+$ define the functions $\Psi_A$ and $\Psi_B$, which allow introducing the Double Parton Distribution Functions and thus expressing the DPI cross section in the familiar factorized form.

\begin{eqnarray}
\int\frac{\psi_A}{l_1^2l_2^2}T_1T_2\frac{\psi_B}{a_1^2a_2^2}d\delta_{-}d\delta_{+}\approx\Bigl(\int d\delta_{-}\frac{\psi_A}{l_1^2l_2^2}\Bigr)\times
\Bigl\{T_1T_2\Bigr\}\times\Bigl(\int d\delta_{+}{\frac{\psi_B}{a_1^2a_2^2}}\Bigr)\equiv\Psi_A\times\Bigl\{T_1T_2\Bigr\}\times\Psi_B
\end{eqnarray}

The case of the $3\to4$ contributions is more elaborate. A partonic process contributing to the DPI amplitude is schematically shown in Fig.4. When looking at Feynman diagrams, one finds that, in the case of all gluons, the number of diagrams to be considered for the 7-gluon amplitude is huge: 2485 at tree level. A standard approach to the problem is therefore impractical. In the zero mass case, tree level amplitudes are nevertheless successfully worked out with the Spinor-Helicity formalism and, in the all-gluon case, tree level amplitudes have been worked out explicitly for any number of external gluons\cite{Britto:2005dg}.

In the spinor-helicity formalism, the tree level amplitude of $n$ gluons with colors $c_1,c_2\dots c_n$, momenta $p_1,p_2\dots p_n$ and helicities $\epsilon_1,\epsilon_2\dots\epsilon_n$, is expressed as 

\begin{eqnarray}\label{eq:M}
{\cal{M}}_n=\sum_{perms'}{\rm Tr}(T^{c_1}T^{c_2}\dots T^{c_n})A(p_1,\epsilon_1;p_2,\epsilon_2;\dots;p_n,\epsilon_n)
\end{eqnarray}

\noindent
where the sum, $perms'$, is over all $non-cyclic$ permutations of 1,2,\dots,$n$ and the $T$'s are the SU(3) generators. The partial amplitudes $A(1,2,...,n)$ are called color ordered amplitudes and satisfy various important properties. In particular each $A(1,2,...,n)$ is a gauge invariant quantity and each different term is incoherent in ${\cal{M}}_n$ to leading order in the number of colors\cite{Mangano:1990by}.
	
The partial amplitudes are expressed in terms of the spinor products:

\begin{eqnarray}
\langle ij\rangle=\sqrt{s_{ij}}\,e^{i\varphi},\qquad[ij]=\sqrt{s_{ij}}\,e^{-i\varphi}
\end{eqnarray}

\noindent
where $s_{ij}=(p_i+p_j)^2$ and $\varphi$ is a phase factor that, in many cases, is not relevant for the final result. All contributions to the 7-gluon amplitudes of interest can be obtained from 4 different color ordered amplitudes\cite{Britto:2004ap}. Each one is expressed by various terms, characterized by singularities in different combinations of the external momenta, where the convention is to define all external momenta in the amplitude as outgoing. The simplest color ordered amplitude is

\begin{eqnarray}
&&A(1^-2^-3^-4^+5^+6^+7^+)\cr
&&\qquad=A(1^-2^-3^-4^+5^+6^+7^+)|_a+A(1^-2^-3^-4^+5^+6^+7^+)|_b+A(1^-2^-3^-4^+5^+6^+7^+)|_c
\end{eqnarray}

As apparent in Fig.4, to contribute to a DPI, a multi-parton amplitude has to be characterized by at least two multi-particle singularities. In the actual case the condition is satisfied by $A|_c$:

\begin{eqnarray}
A(1^-2^-3^-4^+5^+6^+7^+)|_c=-\frac{\langle3|(4+5)(6+7)|1\rangle^3}{P_{345}^2P_{671}^2\langle 34\rangle\langle 45\rangle\langle 6|7+1|2]\langle67\rangle\langle71\rangle\langle5|4+3|2]}
\end{eqnarray}

\begin{wrapfigure}{r}{0.5\textwidth}
  \vspace{-25pt}   
  \begin{center}
    \includegraphics[width=0.45\textwidth]{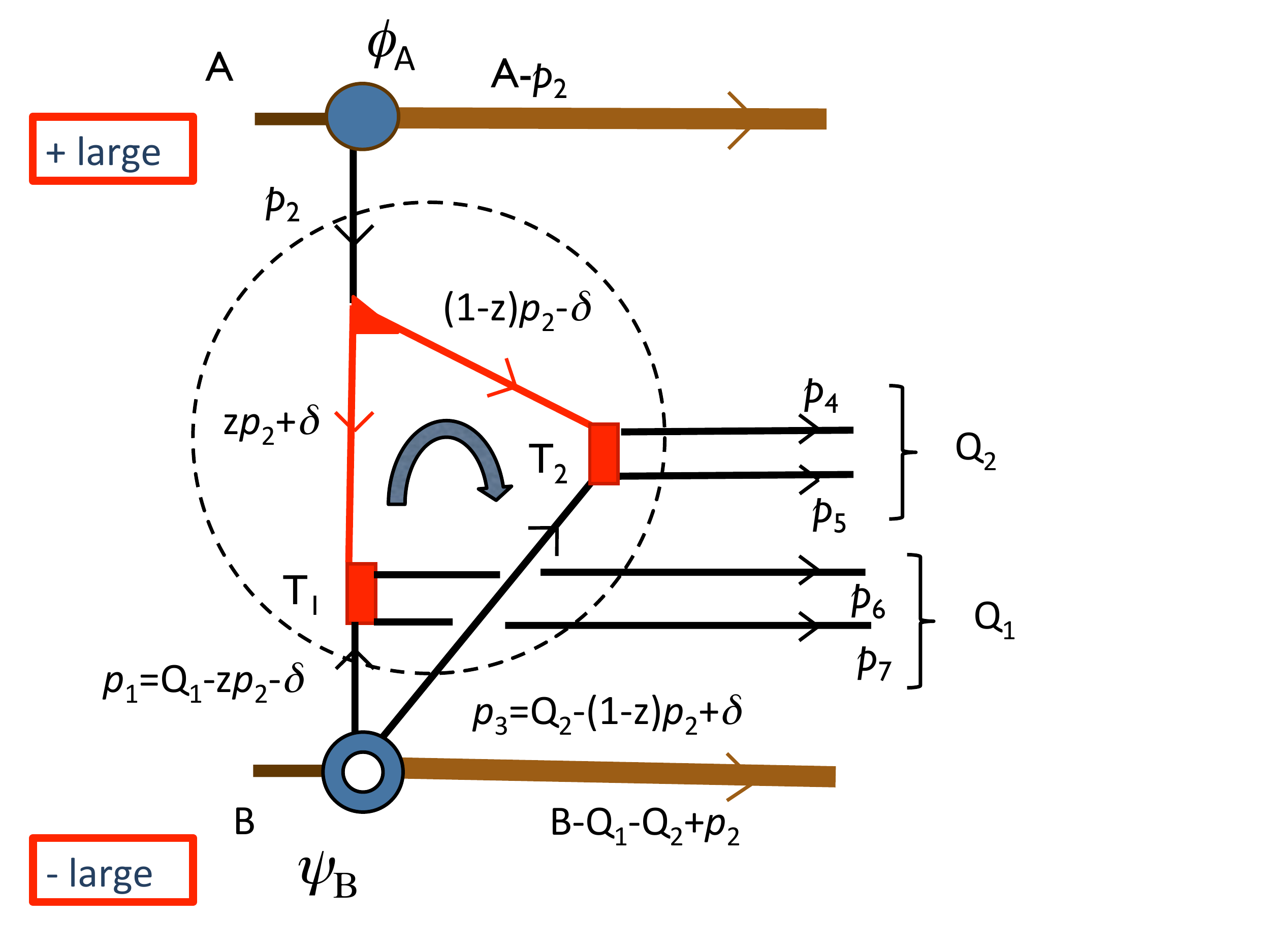}
  \end{center}
  \vspace{-15pt}
  \caption{\small $3\to4$ contributions to the DPI interaction amplitude}
  \vspace{-10pt}
\end{wrapfigure}

Analogously to the $4\to4$ case, to study the effect of the loop integration on the term $A|_c$, one can integrate on the loop integration variables $\delta_+$ and $\delta_-$ (cfr. Fig.4) by keeping into account of the dependence on $\delta_-$ only in the upper part of the loop and of the dependence on $\delta_+$ only in the lower part of the loop. The integration on $\delta_+$ thus defines the function $\Psi_B$ and fixes the values of the fractional momentum components of $P_{345}$ and of $P_{671}$ with respect to $p_2$, actually $z$ and $1-z$, while the integration on $\delta_-$ is estimated with the singularities of $1/(P^2_{345}P^2_{671})$ in $A|_c$:

\begin{eqnarray}
&&P_{345}\equiv -p_3+p_4+p_5\equiv zp_2+\delta,\,\,\,P_{671}\equiv p_6+p_7-p_1\equiv (1-z)p_2-\delta\cr
&& \cr
&&\int d\delta_{-}\frac{1}{P_{345}^2P_{671}^2}=\int \frac{d\delta_{-}}
{\bigl((zp_2+\delta)^2+i\epsilon\bigr)\bigl((1-z)p_2-\delta)^2+i\epsilon\bigr)}=\frac{2\pi i}{\delta_t^2\, p_{2+}}
\end{eqnarray}

The final expression of the amplitude is obtained after integrating on the transverse components $\delta_t$.
Notice that when $\delta_t=0$ the two intermediate gluons, with momenta $P_{345}$ and $P_{671}$, are on mass shell and the process factorizes into the product of a splitting amplitude and of two on shell scattering amplitudes, in such a way that the final state has the characteristic signature of a DPI, namely the transverse momenta compensated pairwise.  

A main point is thus the behavior of the integrand in the limit of small $\delta_t$. At small $\delta_t$ one has:

\begin{eqnarray}
\int A|_c d\delta_-\simeq\frac{\bigl(z\langle32\rangle\langle\delta1\rangle+(1-z)\langle3\delta\rangle\langle21\rangle\bigr)^3[\delta2] }
{\langle34\rangle\langle45\rangle\langle67\rangle\langle71\rangle\langle\delta6\rangle\langle\delta5\rangle}\times\frac{2\pi i}{\delta_t^2 p_{2+}}\to\frac{\rm const.}{\delta_t}\quad{\rm for}\quad\delta_t\to0
\end{eqnarray}

The integration of $A|_c$ on $d^2\delta_t$  thus washes out the singularity at $\delta_t=0$, which implies that the amplitude is {\it not enhanced} in the configuration where the transverse momenta of the two pairs of large $p_t$ partons are compensated pairwise. The contribution of $A|_c$ is thus of the same order of magnitude of the other two contributions $A|_a$ and $A|_b$  to the same gauge invariant color ordered amplitude $A$.

One can show that this property does not hold only for the particular case discussed here. It holds also for all other contributions to the tree level 7-gluon amplitude, characterized by multi-particle singularities.

At tree level, in each color ordered amplitude, one thus finds terms, which can be factorized into a splitting amplitude and two almost on shell  $2\to2$ scattering amplitudes. However these contributions to the color ordered amplitude are {\it not enhanced}, because of the loop integration on the initial momenta, and have to be added coherently with all other contributions (which are of similar magnitude) to the same, gauge invariant, color ordered amplitude. One will then conclude that the $3\to4$ terms cannot contribute in a relevant amount to cross sections, characterized by final state configurations where transverse momenta are compensated pairwise.

\begin{wrapfigure}{r}{0.5\textwidth}
  \vspace{-20pt}
  \begin{center}
    \includegraphics[width=0.48\textwidth]{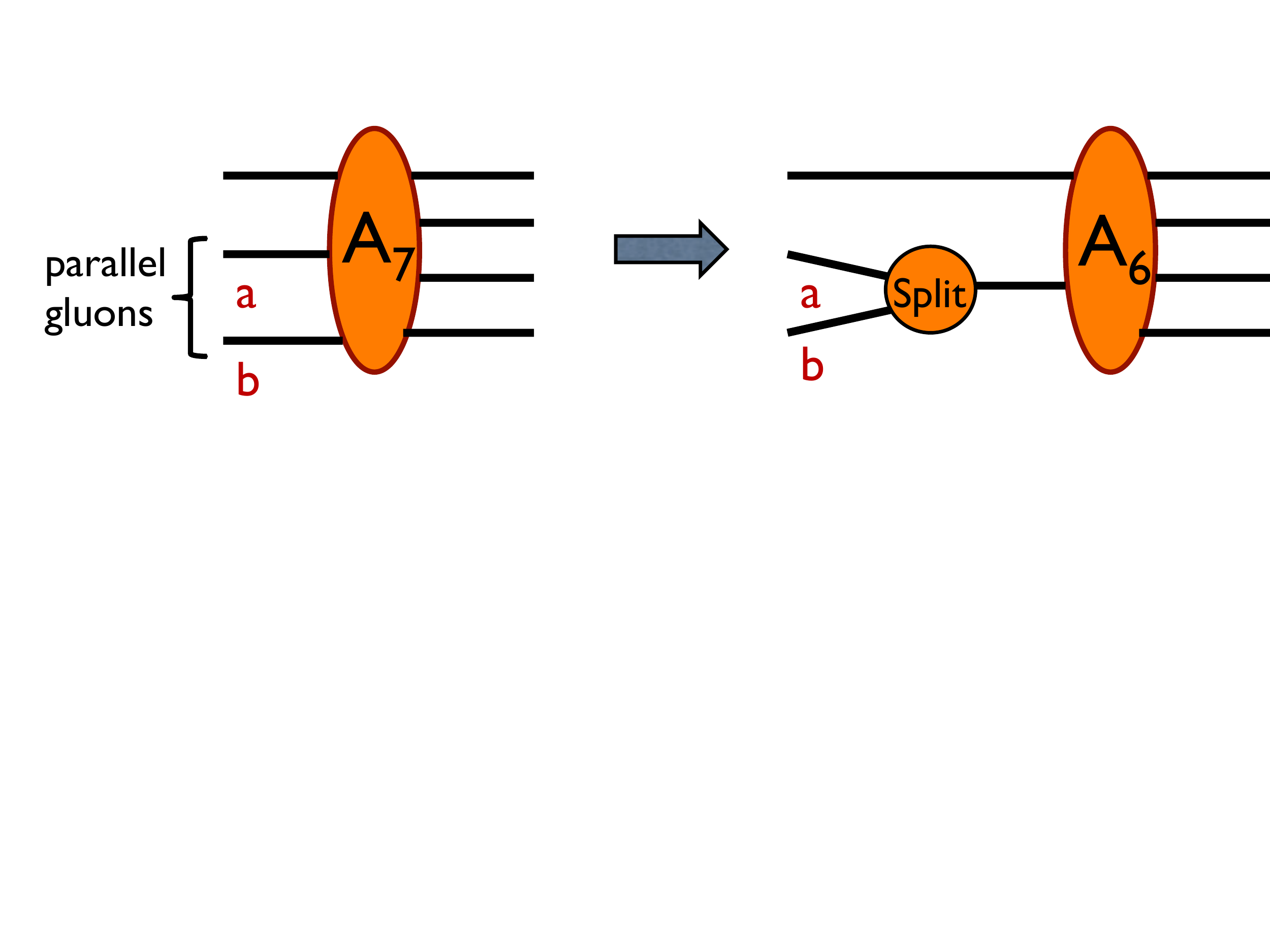}
  \end{center}
  \vspace{-80pt}
  \caption{\small 7-gluon amplitude with two parallel gluons in the initial state}
  \vspace{-5pt}
\end{wrapfigure}

A further remark is that the complete $3\to4$ scattering amplitude results from the sum of all color ordered contributions. In the actual case of interest one has two gluons in the initial state, which are originated by the same hadron and which thus have a rather small relative transverse momentum.
The color ordered terms, where the almost parallel initial state gluons are cyclically-adjacent in the amplitude, are singular in the invariant obtained by the sum of the two almost parallel momenta\cite{Dixon:2013uaa} and therefore give a leading contribution to the amplitude.
A main contribution to the 7-gluon amplitude, in the kinematics considered here, is therefore factorized into a fusion amplitude and a 6-gluon scattering amplitude, the latter with only two gluons in the initial state. 

While initiated by three partons, a main contribution to the cross section is thus effectively given by a $2\to4$, rather than by a $3\to4$ parton process.


\begin{thebibliography}{99}

\bibitem{Paver:1982yp} 
  N.~Paver and D.~Treleani,
  Nuovo Cim.\ A {\bf 70}, 215 (1982).

\bibitem{Blok:2011bu}
  B.~Blok, Y.~Dokshitser, L.~Frankfurt and M.~Strikman,
  Eur.\ Phys.\ J.\ C {\bf 72}, 1963 (2012)
  [arXiv:1106.5533 [hep-ph]].

\bibitem{Diehl:2011yj} 
  M.~Diehl, D.~Ostermeier and A.~Schafer,
  JHEP {\bf 1203}, 089 (2012)
  [arXiv:1111.0910 [hep-ph]].

\bibitem{Britto:2005dg} 
  R.~Britto, B.~Feng, R.~Roiban, M.~Spradlin and A.~Volovich,
  Phys.\ Rev.\ D {\bf 71}, 105017 (2005)
  [hep-th/0503198].

\bibitem{Mangano:1990by} 
  M.~L.~Mangano and S.~J.~Parke,
  Phys.\ Rept.\  {\bf 200}, 301 (1991)
  [hep-th/0509223].

\bibitem{Britto:2004ap} 
  R.~Britto, F.~Cachazo and B.~Feng,
  Nucl.\ Phys.\ B {\bf 715}, 499 (2005)
  [hep-th/0412308].

\bibitem{Dixon:2013uaa} 
  L.~J.~Dixon,
  arXiv:1310.5353 [hep-ph].

\end{thebibliography}
\end{document}